\documentclass[prl,twocolumn,showpacs]{revtex4}
\usepackage{epsfig}
\usepackage{amsmath}

\begin{document}

\title{Dynamics of disordered vortex matter in type II superconductors.}
\author{Golan Bel$^{1}$ and Baruch Rosenstein$^{2,3}$}
\affiliation{$^{1}$Physics Department, Bar-Ilan University, Ramat-Gan 52900, Israel.\\
$^{2}$Electrophysics Department, National Chiao Tung University, Hsinchu
30050, Taiwan, R.O.C.\\ $^{3}$ Department of Condensed Matter Physics, The Weizmann Institute of Science,
Rehovot 76100, Israel.}
\date{\today }

\begin{abstract}
Dynamics of homogeneous moving vortex matter is considered beyond the linear
response. The framework is the time dependent Ginzburg - Landau equation
within the lowest Landau level approximation. Both disorder and thermal
fluctuations are included using the Martin-Siggia-Rose formalism . We
determine the critical current as function of magnetic field and temperature 
$J_{c}\left( B,T\right) $. The surface in the $J-B-T$ space defined by the
function separates between the dissipative moving vortex matter regime (flux
flow) and an amorphous vortex "glass". Both the thermal depinning and the
depinning by a driving force are taken into account. The static
irreversibility line, determined by $J_{c}\left( B,T\right)=0$ is compared
to experiments in layered HTSC and is consistent with the one obtained using the
replica approach. The non-Ohmic I-V curve (in the depinned phase) is
obtained and compared with experiment in layered superconductors and thin
films.
\end{abstract}

\maketitle

In type II superconductors the magnetic field penetrates the sample in a
form of Abrikosov vortices, which interact strongly creating an elastic
"vortex matter". Impurities greatly affect the thermodynamic and especially
dynamic properties of the vortex matter. In high $T_{c}$ superconductors (HTSC)
thermal fluctuations also influence the vortex matter,
either directly by melting the vortex lattice into a liquid or by reducing
the efficiency of the disorder (depinning). As a result of the delicate
interplay between disorder, interactions and thermal fluctuations even the
static $H-T$ phase diagram of HTSC is very complex and is still far from
being reliably determined. Once electric current $J$ is injected into the
sample, the dynamical phase diagram should be drawn in the three dimensional
$T-H-J$ space, which makes the analysis even more complicated. Generally there
are two phases, the pinned phase in which the linear resistivity vanishes
and a phase in which vortices can move due to Lorentz force and thus a
finite resistivity appears. The surface separating the two phases is determined by the critical
current (neglecting very small creep effect) as function of
magnetic field $H$ and temperature $T$. The intersection of the surface with
the $H-T$ plane gives the static irreversibility line.

Theoretically two major simplifications are generally made. In London
approximation (valid far from $H_{c2}$) the vortex matter behaves as an
array of elastic lines \cite{Blater}. An alternative simplification to the
vortex matter, valid far from $H_{c1},$ where the magnetic field is nearly
homogeneous due to overlaps between fields of the\ vortices, is the lowest
Landau level (LLL) approximation \cite{Thouless,Hu}. The original idea of
the vortex glass and the continuous glass transition (GT) was studied in the 
static frustrated XY model \cite{Fisher,Natterman} using RG, variational methods 
and numerical simulations \cite{Hu1}.
It was found that in this model the conductivity exhibits the glass scaling. 
In analogy to
the theory of spin glasses, in this model the replica symmetry is broken
when crossing the GT line. More realistic model, the elastic medium of
interacting line-like objects subject to both the pinning potential and the
thermal bath, was treated using the Gaussian approximation \cite{Giamarchi,Kroshunov} 
and RG \cite{Natterman}. The dynamics in the presence of thermal fluctuations, within the latter model,
can be simulated using the thermal bath Langevin force \cite{Otterlo}.

The irreversibility line (along with other properties) in 2D and 3D disordered vortex matter
was found by applying the replica method to the Ginzburg-Landau (GL) model \cite{Lopatin,Li} 
(supersymmetry was also employed to describe
the effect of columnar defects in layered HTSC \cite{Tesanovich}). The irreversibility
line of YBCO and a 2D organic superconductor were in good agreement with
experiment. Dynamics in the presence of thermal fluctuations and disorder is
generally described using the time dependent Ginzburg - Landau (TDGL) model,
in which the coefficients have random components \cite{Blater}. This model
was studied by Dorsey, Fisher and Huang \cite{DFH} in the homogeneous
(liquid) phase using the dynamic Martin-Siggia-Rose (MSR) approach \cite{MSRSomp}. 
They obtained the irreversibility line and claimed that it is
inconsistent with experiments in YBCO.

It is the purpose of this paper to study the dynamics of vortex matter
within the TDGL model beyond linear response using the dynamical approach of 
\cite{DFH}. We calculate the I-V curve in the homogeneous flux flow phase.
The critical surface in the three dimensional $T-H-J$ space, separating the
pinned and unpinned phases is obtained. The static GT line (zero current) 
coincides with the one obtained using the replica method \cite{Li}. A
relation between the dynamical and the replica methods, which in our mind is
crucial for understanding the nature of any glass transition, is discussed.
Comparison of the irreversibility line, critical current and resistivity
with experimental results in layered superconductors and thin films is made.

Our starting point is the TDGL equation \cite{Larkin} in the presence of
thermal fluctuations on the mesoscopic scale represented by a white noise $%
\zeta $:
\begin{equation}
\frac{{\hbar }^{2}\gamma }{4m^{\ast }}D_{\tau }\psi =-\frac{\delta }{\delta
\psi ^{\ast }}F+\zeta ,  \label{fullTDGL}
\end{equation}
where $m^{\ast }$ is the effective mass of Cooper pair and $\gamma $ is the
inverse diffusion constant. The covariant time derivative is 
$\ D_{\tau}\equiv \frac{\partial }{\partial \tau }+\frac{ie^{\ast }}{\hbar }\Phi$,
where $\Phi $ is the scalar potential describing the driving electric force.
The variance of the thermal noise $\zeta $ determines the temperature via the fluctuation-dissipation
relation:
$\left\langle \zeta \left( x,\tau \right) \zeta ^{\ast }\left( y,\tau
^{\prime }\right) \right\rangle =\delta \left( \tau -\tau ^{\prime }\right)
\delta \left( x-y\right) \frac{{\hbar }^{2}\gamma }{8m^{\ast }}T$. The
static GL free energy including the $\Delta T_{c}$ disorder is: 
\begin{equation}
F=\int d^{3}x\frac{{\hbar }^{2}}{2m^{\ast }}|\overrightarrow{D}\psi
|^{2}-a^{\prime }\left( 1+U\right) |\psi |^{2}+\frac{b^{\prime }}{2}|\psi
|^{4},  \label{feGL}
\end{equation}
where the disorder variance is expressed via dimensionless pinning strength $n$ 
and the coherence length $\xi$:
$\left\langle U\left( x\right) U\left( y\right) \right\rangle =\delta\left( x-y\right) \xi ^{2}n$.
The covariant derivative $\overrightarrow{D}\equiv \vec{\nabla}+\frac{ie^{\ast }}{\hbar c}\vec{A}$
describes the magnetic field, and the coefficients in Eq.(\ref{feGL}) are related to the
coherence length and magnetic penetration depth 
$\lambda $, namely $a^{\prime}\left( T=0\right) =\frac{{\hbar }^{2}}{2\xi ^{2}m^{\ast }}$ and 
$b^{\prime }=\frac{2\pi {\hbar }^{2}\lambda ^{2}e^{\ast 2}}{\xi^{2}c^{2}m^{\ast 2}}$. 
The TDGL equation can be written as 
$\widehat{H}\psi =a^{\prime }U\psi -b^{\prime }|\psi |^{2}\psi +\widetilde{\zeta }$, 
where the (non Hermitian) linear operator is defined by 
$\widehat{H}\equiv \frac{{\hbar }^{2}\gamma }{4m^{\ast }}D_{\tau }-\frac{{\hbar }^{2}}{2m^{\ast
}}D^{2}-a^{\prime }$.
Several assumptions (identical to those used in \cite{Moore} and major parts of \cite{Hu}) are made. 
In strongly type II
superconductors, where $\kappa $ $>>1$, magnetic and electric fields are very
homogeneous, since fields of vortices overlap. Therefore the Maxwell
equations for electromagnetic field are not considered. The Landau gauge\
with vector potential $\vec{A}=(-By,0,0)$ and scalar potential $\Phi =Ey$ is
used. Temperature, current and magnetic field should be close
enough to the dynamical phase
transition line $H_{c2}(T,J)$ in order to apply the GL approach.\newline
\indent In order to perform the averaging over both the thermal fluctuations and
disorder in the dynamical situation, we adapt the MSR formalism \cite{MSRSomp} 
to the present case. The dynamical "partition function" is defined
as a functional integral over the order parameter $\psi$ and an additional
"ghost" field $\phi $ (originating from integral representation of delta function):
\begin{equation}
Z=\int D\psi D\phi \exp \left( -A_{MSR}\left[\psi, \phi, U\right]\right).
\end{equation}
Although the formalism can be applied to both the 3D and the 2D cases, to
simplify the discussion, we consider a 2D superconductor of thickness $L_{z}$.
The MSR dimensionless "action" $A$, with disorder averaged out is:
\begin{widetext}
\begin{equation}
A_{MSR}=-i\frac{L_{z}}{T}\int_{r,,t}\left\{ 
\begin{array}{c}
\left[ \phi _{r,t}^{\ast }\widehat{H}\psi _{r,t}+cc\right] 
+i\frac{{\hbar }^{2}\gamma }{2m^{\ast }}\left\vert \phi _{r,t}\right\vert
^{2}+b^{\prime }|\psi _{r,t}|^{2}\left[ \phi _{r,t}^{\ast }\psi _{r,t}+cc%
\right]  
+irb^{\prime }\int_{s}\left[ \phi _{r,t}^{\ast }\psi _{r,t}+cc\right] \left[
\phi _{r,s}^{\ast }\psi _{r,s}+cc\right] 
\end{array}%
\right\},   \nonumber
\end{equation}
\end{widetext}
where $r=\frac{n}{2\pi ^{2}\sqrt{2Gi}}\frac{\left( 1-t\right) ^{2}}{t}$ and 
$Gi\equiv \frac{1}{2}\left(\frac{2e^{\ast 2}\kappa ^{2}\xi ^{2}T_{c}}{\pi c^{2}\hbar ^{2}L_{z}}\right)^{2}$ 
are the disorder dimensionless parameter and the 2D Ginzburg number respectively.
In high enough magnetic field, $\phi $ and $\psi $ 
can be expanded in a LLL basis (right eigenfunctions of the operator 
$\widehat{H}$ with lowest eigenvalue) 
$\varphi _{k\omega }=\exp [i\left(
\omega t+kx\right) ]\exp \left[ -\frac{b}{2}\left( y/\xi -k\xi
/b+iv/b\right) ^{2}\right]$.
The dimensionless magnetic field and velocity
are $b=B/B_{c2}$ and $v=e^{\ast }\gamma E\xi ^{3}/4\hbar b$
respectively ( $\gamma $ depends on temperature $\gamma =\gamma _{0}\left(1-t\right) ^{-\eta }$,
for example in BCS $\eta =1,$ 
$\gamma _{0}=48\pi \kappa ^{2}\sigma _{n}/c^{2}$, where $\sigma _{n}$ 
is the normal state conductivity \cite{Blater}). 
The LLL basis will be used from now on. 
The model is still a highly nontrivial field theory and we use the Gaussian
effective action approach to treat it. The approximation was applied to the
LLL Ginzburg - Landau model in the absence of electric field (using diagram
resummation) in Ref. \cite{DFH}, leading to identical gap equation. 
Two point Green functions are the correlator 
$C\left( r,t,r^{\prime },t^{\prime
}\right) =\left\langle \psi _{r^{\prime },t^{\prime }}^{\ast }\psi
_{r,t}\right\rangle $, the response function $R\left( r,t,r^{\prime
},t^{\prime }\right) =\left\langle -i\phi _{r^{\prime },t^{\prime }}^{\ast
}\psi _{r,t}\right\rangle \ $ and the auxiliary field correlator $B\left(
r,t,r^{\prime },t^{\prime }\right) =\left\langle \phi _{r^{\prime
},t^{\prime }}^{\ast }\phi _{r,t}\right\rangle $. 
In a homogeneous dynamical phase (stationary flow) \cite{fn},
the correlators time dependence is just on the time difference $t-t^{\prime }$. 
Within LLL we find a solution, which does not depend on momentum 
$k:C\left( k,t,k^{\prime},t^{\prime }\right) =\overline{C}\left( t-t^{\prime }\right) $. 
The variational parameters therefore will be $R_{\omega },$ $C_{\omega }$ and 
$B_{\omega },$ which we write in a matrix form 
$g=\left( \begin{array}{cc}C & iR \\ -iR^{\ast } & B\end{array}\right)$.

The Gaussian effective action is:
\begin{eqnarray*}
\mathcal{A} &\propto &\int_{\omega }\left[ Tr\log g^{-1}\ +Tr\left(
g_{0}^{-1}g\right) \right] +\ \frac{\theta }{2\pi }\int\limits_{\omega
,\lambda }\ C_{\lambda }\left( R_{\omega }-R_{\omega }^{\ast }\right) \\
&&+\ \frac{r\theta }{2}\int_{\omega }\left[ 2B_{\omega }C_{\omega
}-R_{\omega }^{\ast 2}-R_{\omega }^{2}\right] ,
\end{eqnarray*}
where $\theta \equiv 4\pi tb\sqrt{2Gi}$, $a_{h}=-\left( 1-t-b-v^{2}\right) /2$
and
\[
\ g_{0}^{-1}=\frac{L_{z}\xi \hbar ^{2}e^{v^{2}/b}}{2\pi ^{1/2}m^{\ast
}Tb^{3/2}}\left( 
\begin{array}{cc}
0 & -\omega \gamma \xi ^{2}/2-2ia_{h} \\ 
\omega \gamma \xi ^{2}/2-2ia_{h} & \gamma \xi ^{2}%
\end{array}%
\right) . 
\]
The corresponding "gap equations" are
\begin{equation}
\left( 2a_{h}+\theta C\left( t=0\right) +i\omega \gamma \xi ^{2}/2\right)
R_{\omega }-r\theta R_{\omega }^{2}=1  \label{R}
\end{equation}
\begin{equation}
C_{\omega }=\frac{\gamma \xi ^{2}\left\vert R_{\omega }\right\vert ^{2}}{%
1-r\theta \left\vert R_{\omega }\right\vert ^{2}};B_{\omega }=0.  \label{C}
\end{equation}
For $v=0$ the equations are consistent with those of Ref. \cite{DFH}. 
For $\omega =0,$ one obtains
\begin{equation}
R_{\omega =0}=\frac{-a_{h}+\sqrt{a_{h}^{2}+\theta \left( 1-r\right) }}{%
\theta \left( 1-r\right) }
\label{R0}
\end{equation}
Eqs. (\ref{C},\ref{R0}) are consistent with the static ($t=0$ and $v=0$) correlator calculated
using the replica method \cite{Li}. The solution for arbitrary $\omega $ is:
\[
R_{\omega }=\frac{a_{v}+i\omega \gamma \xi ^{2}/2-\sqrt{\left( a_{v}+i\omega
\gamma \xi ^{2}/2\right) ^{2}-4r\theta }}{2r\theta }
\]
where
\[
a_{v}=\frac{a_{h}\left( 1-2r\right) +\sqrt{a_{h}^{2}+\theta \left(
1-r\right) }}{\left( 1-r\right) }.
\]
\begin{figure}[tbp]
\begin{center}
\epsfxsize=70mm \epsfbox{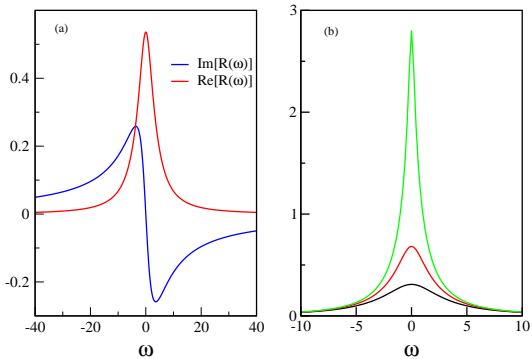}
\end{center}
\caption{ In (a) the imaginary and real
parts of the response function are shown for arbitrary values of the
parameters. The correlation function has a peak at $\protect\omega =0.$ In
(b) it is shown qualitatively how the correlation function at zero frequency
diverges as the parameters approach the critical values. }
\label{fig1}
\end{figure}
The correlator and the response
functions versus frequency are plotted in Fig. \ref{fig1}. The
correlator decreases as $1/\omega ^{2}$ at large frequencies. The structure
of the singularities in the complex $\omega $ plane is as follows. In
addition to cuts there is a segment of singularities along the positive
imaginary axis between $\omega _{\min }^{\max }$ 
$=2\gamma ^{-1}\xi^{-2}\left( a_{v}\pm \sqrt{4r\theta }\right) $. 
This range of frequencies
corresponds to a range of relaxation times. Asymptotically the long time
dependence of the correlator is exponential, 
$C\left( t-t^{\prime }\right)\propto \exp \left[ -\frac{t-t^{\prime }}{\tau _{\max }}\right] $. 
The dominant time scale for relaxation is the longest one, given by $\tau _{\max }=\frac{1}{\omega _{\min }}$ 
which diverges as one approaches the critical surface determined next.

There are two dynamical phases of the system in the tree dimensional
external parameter space $(T,H,J).$ It is more convenient to use instead
dimensionless scaled parameters $(t,b,v).$ The critical surface is defined
as a set of values of the parameters for which the correlator $C_{\omega }$
at $\omega =0$ diverges. We will argue later that below this surface the
superconductor acquires certain "glassy" properties. According to Eq. (\ref{C}), 
the correlator at zero frequency diverges when the denominator vanishes,
namely
\begin{equation}
1-r\theta \left\vert R_{\omega =0}^{g}\right\vert ^{2}=0.
\label{criticality}
\end{equation}
Using the response function Eq. (\ref{R0}), one obtains the critical
surface
\begin{equation}
a_{T}^{g}\left( v\right) =4r^{1/2}-2r^{-1/2},  \label{aTglass}
\end{equation}
where the LLL scaled temperature is 
$a_{T}\left( v\right) \equiv 4a_{h}\theta ^{-1/2}$. 
In terms of the original dimensionless parameters 
$t,b,v$ it takes the form:
\begin{equation}
1-t-b-v^{2}=\theta ^{1/2}\left( r^{-1/2}-2r^{1/2}\right) .  \label{gl}
\end{equation}
\begin{figure}[tbp]
\begin{center}
\epsfxsize=70mm \epsfbox{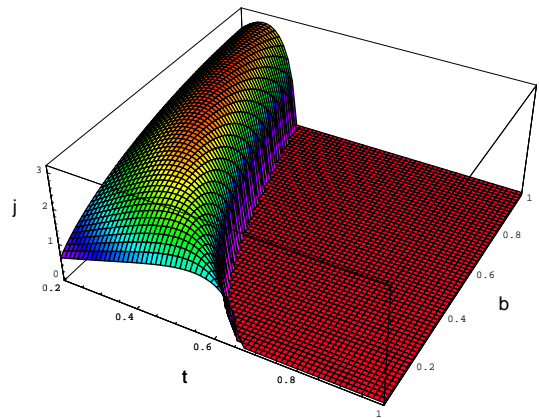}
\end{center}
\caption{ The Irreversibility surface in
the dimensionless $j,t,b,$ parameters space.}
\label{fig2}
\end{figure}
The static irreversibility line ($v=0$), which is the intersection of the
dynamical glass transition surface (Fig. \ref{fig2}) with the $H-T$ plane,
has been observed in great variety of type II superconductors in magnetic
fields. Within the range of applicability of the LLL approximation the
static irreversibility line in both 2D and 3D was already obtained using
both the replica method \cite{Li} and the dynamical approach \cite{DFH}.
The original impression at \cite{DFH} was that the line is inconsistent with
experiments in thin films of YBCO, however recently the irreversibility
line in organic superconductor $\kappa $ type $BEDT-TTF$, was found to be
in good agreement with the experiment \cite{Li}.  
In Fig. \ref{fig3} we compare the irreversibility line of BSCCO in high magnetic fields \cite{Ando}
with Eq. (\ref{gl}) for $J=v=0$ and the following values of the material parameters: 
$H_{c2}=195T,T_{c}=93K,$ $Gi=0.00044,n=0.005$.
\begin{figure}[tbp]
\begin{center}
\epsfxsize=70mm \epsfbox{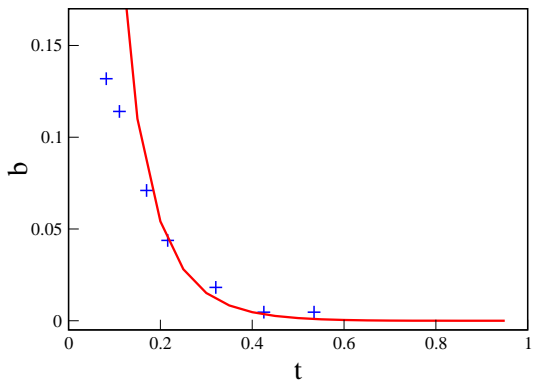}
\end{center}
\caption{ The glass line in the b-t
plane. The pluses correspond to experimental data of \cite{Ando},
and the solid curve corresponds to the analytical formula Eq. (\ref{gl}), where $v=0$.}
\label{fig3}
\end{figure}
One would like to parametrize the surface in terms of the supercurrent
density \cite{Moore} rather than in terms of the flux velocity $v$. It
provides the nonlinear I-V curve:
\begin{eqnarray}
J_{S} &=&\frac{2e^{\ast }vTb}{L_{z}\xi \hbar }R_{\omega =0}  \nonumber \\
&=&\frac{\hbar c^{2}v}{\xi e^{\ast }\lambda ^{2}}\theta ^{1/2}\frac{%
-a_{T}\left( v\right) +\sqrt{a_{T}^{2}\left( v\right) +16\left( 1-r\right) }%
}{16\left( 1-r\right) }.
\label{js}
\end{eqnarray}
Note that Eq. (\ref{js}) is valid only for $J_{S}>J_{c}$.
The critical current according to Eq. (\ref{gl}) is:
\begin{equation}
J_{c}=\frac{\hbar c^{2}\theta ^{1/2}}{4e^{\ast }\lambda ^{2}\xi r^{1/2}}%
\left[ 1-t-b+2\theta ^{1/2}\left( 2r^{1/2}-r^{-1/2}\right) \right] ^{1/2}.
\label{jc}
\end{equation}
The critical current of the $MoGe$ films \cite{Kapitulnik} as function of
temperature compares qualitatively well with Eq. (\ref{jc}) in the region
beyond the peak effect (namely in the homogeneous phase).

\begin{figure}[tbp]
\begin{center}
\epsfxsize=70mm \epsfbox{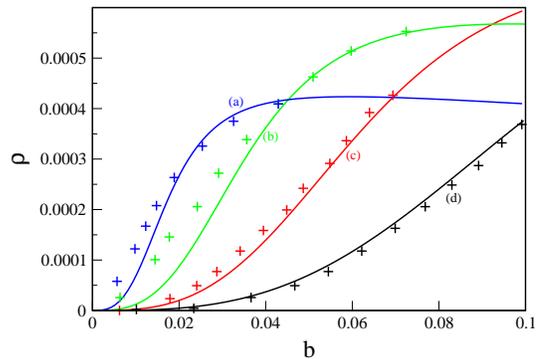}
\end{center}
\caption{ Resistivity as function of the magnetic field. 
The solid curves show the analytical results (Eq. (\ref{sigma})), 
while the pluses correspond to experimental results of \cite{Ando}.}
\label{fig4}
\end{figure}

In the linear response limit Eq. (\ref{js}) determines the conductivity due
to Cooper pairs:
\begin{equation}
\sigma _{S}=\sigma _{n}\frac{3\pi ^{3/2}}{2}\left( 2Gi\right) ^{1/4}\sqrt{%
\frac{t}{b}}\frac{a_{T}\left( 0\right) -\sqrt{a_{T}^{2}\left( 0\right)
+16\left( 1-r\right) }}{1-r}.  \label{sigma}
\end{equation}
This expression, valid for non-zero electric field, 
has a finite limit when one approaches the critical surface. Within
the vortex glass theory \cite{Fisher} the conductivity diverges, and in 
\cite{DFH} it was argued that higher order corrections to the Gaussian
approximation lead to this divergence. We were unable to confirm that and
believe that the exponentially small creep may appear when instanton effects
are taken into account and higher Landau levels are added. Comparison of the
resistivity in BSCCO \cite{Ando} with the one obtained from Eq. (\ref{sigma}) 
(taking into account the normal part of the conductivity) is shown in Fig. \ref{fig4}.
We conclude by discussion of the transition to the glassy state. When the
distance from the critical surface in the parameter space $\Delta
=a_{T}\left( v\right) -a_{T}^{g}\left( v\right) $ approaches zero, certain
physical quantities diverge powerwise. For example, the relaxation time
diverges as $\tau _{\max }\propto \left( a_{T}-a_{T}^{g}\right) ^{-2}$ and $%
C\left( \omega =0\right) \propto \left( a_{T}-a_{T}^{g}\right) ^{-1}$. Note
however that the static correlator $C\left( t=t^{\prime }\right) $
(proportional to the magnetization within LLL \cite{Moore}) does not diverge
at the critical surface. Thus it can not serve as an order parameter for the
"glass" transition, even in the static limit. The common wisdom is that
replica symmetry is broken in the
glass (either via steps or via
hierarchical continuous process) as in
most of the spin glasses theories \cite{Dottsenko}. The replica method
applied to the static LLL model with $\Delta T_{c}$ disorder within Gaussian
approximation \cite{Li} indicates that there is no replica symmetry breaking
in the homogeneous phase. However the Edwards -Anderson parameter vanish
above the GT, while is nonzero below it. This is in agreement with the
original approach to the glass transition of EA (see \cite{Bouchaud} for a
discussion). The results obtained here demonstrate criticality in
this case.\newline
\indent To summarize, using the dynamical approach we obtained the dynamical
critical surface separating the liquid and glass phases. In particular the
static irreversibility line was obtained, and shown to be in a good
agreement with experiment in BSCCO. The resistivity is also found to be in a
good agreement with experimental results. I-V curve and critical current are
calculated beyond the linear response limit using the dynamical approach at
finite electric field.

\textbf{Acknowledgements}

We would like to thank V. Vinokur, A. Koshelev, A. Dorsey and D. P. Li for
fruitful discussions. We would also like to thank P. Kes, N. Kokubo and E. Zeldov
for very helpful discussions of the experimental point of view on the
questions under consideration in this paper.
B. R. acknowlegdes the support of Albert Einstein Minerva Center for
Theoretical Physics at Weizmann Institute, NSC of Taiwan grant 94-2112-M-009-009
and hospitality at Bar Ilan University. G. B. Thanks the NCTS in Taiwan for hospitality at the beginning of
the project.


\begin{thebibliography}{99}
\bibitem{fn} Recent work on statics using the replica method \cite{Li} showed that
lateral correlations (which distinguish between a homogeneous and
crystalline states) are not important for the location of the
irreversibility line, thus we study a simpler homogeneous situation. Note
that the irreversibility line is different from the melting line separating the
liquid and crystalline phases of the vortex matter \cite{Zeldov}.
\end{thebibliography}
\end{document}